# Energy conditions in the $f(R, L, T)$ theory of gravity

Simran Arora[1] [a], P.H.R.S. Moraes[2] [b], and P.K. Sahoo[1] [c]

[1] Department of Mathematics, Birla Institute of Technology and Science-Pilani, Hyderabad Campus, Hyderabad 500078, India
[2] Laboratório de Física Teórica e Computacional (LFTC), Universidade Cidade de São Paulo (UNICID) - Rua Galvão Bueno 868, 01506-000 São Paulo, Brazil

**Abstract.** We construct the energy conditions for the recently proposed $f(R, L, T)$ gravity theory, for which $f$ is a generic function of the Ricci scalar $R$, matter lagrangian density $L$ and trace of the energy-momentum tensor $T$. We analyse two different forms for the $f(R, L, T)$ function within the framework of the Friedmann-Lemâitre-Robertson-Walker universe. We constrain the model parameters from the energy conditions. This approach allows us to assess the feasibility of specific forms of the $f(R, L, T)$ gravity.

## 1 Introduction

Since the discovery of the acceleration of the universe expansion in the late 90's of the last century [1, 2], theoretical physicists struggle to discover what mechanism is behind such an outstanding dynamical phenomenon. In principle, a simple constant term, namely *cosmological constant* $\Lambda$, in Einstein's field equations of General Relativity (GR), could provide the late acceleration dynamics. However, this approach is plagued by the so-called *cosmological constant problem*, which is the extraordinarily high discrepancy between the observational [1, 2] and theoretical values of $\Lambda$ [3].

This shortcoming has led to the consideration of the possibility that what really drives the universe expansion to accelerate is an effect "hidden" in a wider formalism of gravity. We already faced this kind of paradigm shift when GR theory predicted the observed "discrepancies" in Mercury's orbit [4], which were not present in Newtonian formalism. We might be facing a new paradigm shift.

An extended gravity formalism must be able to recover GR in certain regimes, as the Solar System, but while the standard cosmological model, based on GR, explains the cosmic acceleration through $\Lambda$, an extended gravity cosmological model should explain it from extra terms appearing in the field equations of such theories [5].

Nowadays, we have numerous extended formalisms of gravity, such as $f(R)$ theories [6, 7], $f(R, L)$ gravity [8, 9], $f(R, T)$ gravity [10], $f(\mathcal{T})$ gravity [11, 12], $f(Q)$ gravity [13, 14], $f(Q, T)$ gravity [15, 16, 17, 18, 19], among others, in which $f$ stands for a generic function of the argument(s), and $R$ stands for the Ricci scalar, $L$ for the matter lagrangian density, $T$ for the trace of the energy-momentum tensor, $\mathcal{T}$ for the torsion scalar and $Q$ for the non-metricity scalar.

It is clear the necessity of distinguishing among these models and searching for the valid ones within them. For these purposes, a well-known tool is cosmography [20, 21, 22], which is not in the scope of the present article.

Here, we are going to use the energy conditions in order to constrain a recently proposed gravity theory, called $f(R, L, T)$ theory [23]. The energy conditions refer to physical restrictions imposed to the distribution of matter and energy in space-time. They represent paths to implement the positiveness of the energy-momentum tensor, the attractive nature of gravity and the geodesic structure of space-time. They have been discussed in the literature for a long time, as one can check [24, 25], among others. Some of the references above, namely, [8, 18], are, indeed, energy conditions applications in extended gravity.

An exemplary contribution to the field of energy conditions applications was presented in Reference [26], in which the energy conditions were applied to the $f(Q)$ gravity theory. An important result of [26] was obtained for a well-known functional form of the $f(Q)$ function, namely $f(Q) = Q + mQ^n$, with $m$ and $n$ being free parameters. The constraints obtained from energy conditions application to such a model were $m \leq -1$ and $0.9 \leq n \leq 2$.

In the present article, we apply the energy conditions to the very recently developed $f(R, L, T)$ gravity theory. For this purpose, we introduce the outline of $f(R, L, T)$ theory in Section 2. In Section 3, we discuss the energy condition in the context of GR. We obtain the inequalities corresponding to the energy conditions in $f(R, L, T)$ theory in Section 4. In Section 5, we formulate the expressions for the energy conditions in terms of cosmological parameters and discuss the energy bounds for the two specific $f(R, L, T)$ models. In the last section,

[a] *email:* dawrasimran27@gmail.com
[b] *email:*moraes.phrs@gmail.com
[c] *email:*pksahoo@hyderabad.bits-pilani.ac.in

we summarize our results and present our concluding remarks.

A brief outline of our results is presented below. We consider two functional forms for the $f(R,L,T)$ function, namely $R+\gamma T+\lambda L$ and $R^n+\alpha LT+\beta$, with $\gamma, \lambda, \alpha$ and $\beta$ being free parameters to be constrained by the energy conditions. For the first of these models, we obtain that in order to have an accelerated expansion for the universe, the parameter $\gamma$ must be positive and $\lambda > -16\pi - 2\gamma$. This is a strong constraint that should be respected within further applications of this model. For the second model, the constraints are not strong, but maximum values for $\alpha$ and $\beta$ are likewise derived.

## 2 The $f(R,L,T)$ theory of gravity

Haghani and Harko proposed in [23] a generalization of two gravity theories that allow to non-minimal coupling between geometry and matter, namely $f(R,L)$ and $f(R,T)$ theories (check, respectively, [27] and [10]), the so-called $f(R,L,T)$ theory of gravity.

The $f(R,L,T)$ gravity formalism starts from the action

$$\mathcal{S} = \frac{1}{16\pi}\int d^4x \sqrt{-g} f(R,L,T) + \int d^4x \sqrt{-g} L, \quad (1)$$

in which $g$ stands for the metric determinant.

The variation of the above action with respect to the metric tensor $g_{\mu\nu}$ yields the following field equations of $f(R,L,T)$ gravity

$$(R_{\mu\nu} + g_{\mu\nu}\Box - \nabla_\mu\nabla_\nu)f_R - \frac{1}{2}f g_{\mu\nu} = 8\pi T_{\mu\nu} +$$
$$\frac{1}{2}(f_L + 2f_T)(T_{\mu\nu} - L g_{\mu\nu}) + f_T \tau_{\mu\nu}, \quad (2)$$

in which $R_{\mu\nu}$ is the Ricci tensor, $g_{\mu\nu}$ is the metric tensor, $\Box = g^{\mu\nu}\nabla_\mu\nabla_\nu$, $T_{\mu\nu}$ is the energy-momentum tensor, $f_R \equiv \partial f/\partial R$, $f_L \equiv \partial f/\partial L$, $f_T \equiv \partial f/\partial T$, and $\tau_{\mu\nu} \equiv 2g^{\alpha\beta}\partial^2 L/(\partial g^{\mu\nu}\partial g^{\alpha\beta})$.

By taking the trace of (2), we obtain

$$(R+3\Box)f_R - 2f = 8\pi T + (f_L + 2f_T)\frac{T-4L}{2} + f_T\tau. (3)$$

It is evident that the above field equations (2) recover the metric formalism field equations of the $f(R)$ theory for the functional choice $f(R,L,T) = f(R)$. Moreover, if $f(R,L,T) = f(R,L)$, it leads to the field equations of $f(R,L)$ theory and an analogous scenario arises in the case of $f(R,L,T) = f(R,T)$. Furthermore, when $f(R,L,T) = R$, we obtain the standard field equations of GR.

From Equation (2), we see that by taking its covariant derivative is likely to yield a non-null result for the covariant derivative of the energy-momentum tensor. This feature is commonly seen in gravity theories that allow for the coupling between geometry and matter, such as the aforementioned $f(R,L)$ and $f(R,T)$ theories. Effectively, this non-conservation of the energy-momentum tensor would be reflected in matter creation during the evolution of the universe, a subject profoundly investigated for these theories in [28]. Anyhow, it is important to mention that a mechanism to evade the energy-momentum non-conservation can be seen in [29], in which the authors took the equation for $\nabla_\mu T^{\mu\nu}$ in the $f(R,T)$ formalism, forced $\nabla_\mu T^{\mu\nu}$ to be null and obtained the referred solution for the $f(R,T)$ functional. The same can be made in the present $f(R,L,T)$ gravity in a future work.

## 3 The energy conditions

An important aspect of the approach that we shall employ to deduce the energy conditions for the $f(R,L,T)$ theory can be adopted from the strategy used to establish the null energy condition (NEC), strong energy condition (SEC), weak energy condition (WEC) and dominant energy condition (DEC) within GR framework. To ensure accuracy, we will begin by providing a concise overview of the methodology used to address these energy conditions in GR [30]. The Raychaudhuri equation, together with the condition of attractive gravity, serves as the fundamental source of the energy conditions [31]. For this, let $u^\mu$ be the tangent vector field in case of a congruence of timeline geodesics. The Raychaudhuri equation is

$$\frac{d\theta}{d\tilde{\tau}} = -\frac{1}{3}\theta^2 - \sigma_{\mu\nu}\sigma^{\mu\nu} + \omega_{\mu\nu}\omega^{\mu\nu} - R_{\mu\nu}u^\mu u^\nu, \quad (4)$$

where $\tilde{\tau}$, $\theta$, $\sigma_{\mu\nu}$ and $\omega_{\mu\nu}$ are the conformal time, expansion parameter, shear and rotation associated with the congruence, respectively. The equation governing the expansion of a congruence of null geodesics, which is determined by a vector field $k^\mu$, exhibits a similar structure to the Raychaudhuri equation.

Since GR field equations relate $R_{\mu\nu}$ to $T_{\mu\nu}$, the combination of Einstein's and Raychaudhuri's equations can be used to restrict the energy-momentum tensor on physical grounds. Since the shear is a spatial tensor, one has $\sigma^2 = \sigma_{\mu\nu}\sigma^{\mu\nu} \geq 0$. The condition for attractive gravity, i.e., $d\theta/d\tilde{\tau} < 0$, for any hypersurface orthogonal congruences ($\omega_{\mu\nu} = 0$), reduces to

$$R_{\mu\nu}u^\mu u^\nu \geq 0, \quad and \quad R_{\mu\nu}k^\mu k^\nu \geq 0. \quad (5)$$

Henceforth, we obtain the following conditions

$$R_{\mu\nu}u^\mu u^\nu = \left(T_{\mu\nu} - \frac{T}{2}g_{\mu\nu}\right)u^\mu u^\nu \geq 0, \quad (6)$$

$$R_{\mu\nu}k^\mu k^\nu = T_{\mu\nu}k^\mu k^\nu \geq 0. \quad (7)$$

Equations (6) and (7) are nothing but SEC and NEC, respectively, in a coordinate invariant way in terms of $T_{\mu\nu}$ and vector fields character. Note that for a perfect fluid of density $\rho$ and pressure $p$, Equations (6) and (7) reduce to $\rho + 3p \geq 0$ and $\rho + p \geq 0$, respectively.

Another energy condition, namely WEC, mandates the positivity of the energy density for any observer at any point, indicating $\rho \geq 0$ and $\rho + p \geq 0$. The requirement that energy cannot propagate faster than the speed of light gives rise to an additional condition, namely $\rho \pm p \geq 0$, along with the condition $\rho \geq 0$, which is called DEC.

## 4 Energy conditions in the $f(R, L, T)$ gravity

Before constructing the energy conditions in the $f(R, L, T)$ gravity, let us briefly present some insightful discussions in the literature about energy conditions in extended gravity theories.

In the realm of modified gravity theories, the generalized field equations can often be expressed in the following form [32]

$$g(\Psi^i)(G_{\mu\nu} + H_{\mu\nu}) = 8\pi G T_{\mu\nu}. \tag{8}$$

In this context, $H_{\mu\nu}$ is an extra geometric term relative to GR that encompasses the geometric modifications introduced by the modified theory. The term $g(\Psi^i)$ is a factor that adjusts the coupling with the matter fields in $T_{\mu\nu}$, in which $\Psi^i$ generally denotes curvature invariants or other gravitational fields influencing the dynamics. We recover GR for $H_{\mu\nu} = 0$ and $g(\Psi^i) = 1$.

Since $H_{\mu\nu}$ is a geometric quantity, meaning it can be expressed through geometric invariants or scalar fields distinct from ordinary matter fields, imposing a specific energy condition on $T_{\mu\nu}$ affects the combination of $G_{\mu\nu}$ and $H_{\mu\nu}$ rather than just the Einstein tensor. Therefore, unlike the GR case [33], we can no longer derive a straightforward geometric implication. In the literature, $H_{\mu\nu}$ is typically treated as a correction to the energy-momentum tensor, leading to the interpretation that energy conditions involve satisfying a specific inequality using the combined quantity $T^{\mu\nu}_{eff} = T^{\mu\nu}/g - H^{\mu\nu}$. Consequently, it can be misleading to directly associate this effective energy-momentum tensor with the energy conditions, as they are influenced not solely by $T_{\mu\nu}$ but also by the geometric quantity $H_{\mu\nu}$.

However, if the modified theory of gravity being considered allows for an equivalent description through an appropriate conformal transformation, it becomes justified to associate the transformed $H_{\mu\nu}$ with the redefined $T_{\mu\nu}$ in the conformally transformed Einstein frame (check also Reference [34]).

The energy conditions were initially formulated within the framework of GR. However, it is possible to reinterpret these conditions directly by introducing new effective pressure and adjusting the energy density accordingly. More specifically, it is possible to recast alternative theories of gravity so that they may be dealt under standard energy conditions.

The analysis done in [32, 35, 36], for any generalized extended theory of gravity, results in the inequality $g(\Psi^i)(R_{\mu\nu} + H_{\mu\nu} - \frac{1}{2}g_{\mu\nu}H)k^\mu k^\nu \geq 0$, which does not necessarily imply $R_{\mu\nu}k^\mu k^\nu \geq 0$. It is noted that one cannot conclude that the attractive nature of gravity is equivalent to the satisfaction of SEC in the particular modified theory under consideration. However, as it was mentioned above, if the modified gravity theory allows an equivalent description upon an appropriate conformal transformation, it then becomes justified to associate the transformed $H_{\mu\nu}$ to the redefined $T_{\mu\nu}$ in the conformally transformed Einstein frame, where matter and geometrical quantities can be formally dealt exactly such as in GR.

We can rewrite Eq.(2) as follows

$$G_{\mu\nu} = R_{\mu\nu} - \frac{1}{2}g_{\mu\nu}R = 8\pi T^{eff}_{\mu\nu}, \tag{9}$$

in which the effective energy-momentum tensor $T^{eff}_{\mu\nu}$ is

$$T^{eff}_{\mu\nu} \equiv \frac{1}{8\pi f_R} \left[ \frac{1}{2}g_{\mu\nu}(f - f_R R) - (g_{\mu\nu}\Box - \nabla_\mu \nabla_\nu) f_R \right.$$
$$\left. + 8\pi T_{\mu\nu} + \frac{1}{2}(f_L + 2f_T)(T_{\mu\nu} - Lg_{\mu\nu}) + f_T \tau_{\mu\nu} \right]. \tag{10}$$

Contracting Equation (10), we obtain

$$T^{eff} = \frac{1}{8\pi f_R} \left[ 2(f - f_R R) - 3\Box f_R + 8\pi T \right.$$
$$\left. + \frac{1}{2}(f_L + 2f_T)(T - 4L) + f_T \tau \right]. \tag{11}$$

Following what was done for the $f(R, T, R_{\mu\nu}T^{\mu\nu})$ gravity in [37], the attractive nature of gravity needs to satisfy the following additional constraint

$$\frac{8\pi + \frac{1}{2}(f_L + 2f_T)}{8\pi f_R} > 0, \tag{12}$$

which does not depend on the conditions derived from the Raychaudhuri equation.

Let us take the Friedmann-Lemaître-Robertson-Walker (FLRW) metric,

$$ds^2 = -dt^2 + a(t)^2 \left( dx^2 + dy^2 + dz^2 \right), \tag{13}$$

in which $a(t)$ is the scale factor. From (13), we obtain $R = 6(2H^2 + \dot{H})$, in which the Hubble parameter $H = \dot{a}/a$.

By using Equation (6), the SEC can be given as $T^{eff}_{\mu\nu}u^\mu u^\nu + \frac{1}{2}T^{eff} \geq 0$, where we used $g_{\mu\nu}u^\mu u^\nu = -1$. Taking the energy-momentum tensor of a perfect fluid, that is, $T_{\mu\nu} = (\rho + p)u_\mu u_\nu + pg_{\mu\nu}$, and considering the condition for SEC in $f(R, L, T)$ theory, yields

$$\frac{1}{8\pi f_R} \left[ (\rho + 3p)\left(8\pi + \frac{1}{2}(f_L + 2f_T)\right) + (f - f_R R) \right. \tag{14}$$
$$\left. + 3\left(f_{RRR}\dot{R}^2 + f_{RR}\ddot{R} + H f_{RR}\dot{R}\right) - (f_L + 2f_T)L \right] \geq 0,$$

where the dot denotes cosmic time differentiation. We neglect the terms involving the second derivative of $L$ with respect to $g_{\mu\nu}$. As we are dealing with perfect fluid, the matter lagrangian can either be $L = p$ or $L = -\rho$, which makes it obvious to ignore such a term.

The NEC in $f(R, L, T)$ gravity can be expressed as

$$\frac{1}{8\pi f_R}\left[(\rho+p)\left(8\pi+\frac{1}{2}(f_L+2f_T)\right)+ \left(f_{RRR}\dot{R}^2+f_{RR}\ddot{R}\right)\right] \geq 0. \quad (15)$$

Note that the above expressions for the SEC and NEC are directly derived from the Raychaudhuri equation. Using Equations (10) and (13), the effective energy density and the effective pressure can be derived as

$$\rho^{eff} = \frac{1}{8\pi f_R}\left[(8\pi+\frac{1}{2}(f_L+2f_T))\rho - \frac{1}{2}(f-f_R R)\right.$$
$$\left. -3Hf_{RR}\dot{R}+\frac{1}{2}(f_L+2f_T)L\right] \geq 0, \quad (16)$$

$$p^{eff} = \frac{1}{8\pi f_R}\left[(8\pi+\frac{1}{2}(f_L+2f_T))p + \frac{1}{2}(f-f_R R)+ \right. \quad (17)$$
$$\left. f_{RRR}\dot{R}^2+f_{RR}\ddot{R}+2Hf_{RR}\dot{R}-\frac{1}{2}(f_L+2f_T)L\right] \geq 0.$$

Then, the corresponding DEC and WEC in $f(R, L, T)$ gravity can be respectively given as

$$\frac{1}{8\pi f_R}\left[(8\pi+\frac{1}{2}(f_L+2f_T))(\rho-p)-(f-f_R R)\right. \quad (18)$$
$$\left. -f_{RRR}\dot{R}^2-f_{RR}\ddot{R}-5Hf_{RR}\dot{R}+(f_L+2f_T)L\right] \geq 0,$$

$$\frac{1}{8\pi f_R}\left[(8\pi+\frac{1}{2}(f_L+2f_T))\rho-\frac{1}{2}(f-f_R R)\right. \quad (19)$$
$$\left. -3Hf_{RR}\dot{R}+\frac{1}{2}(f_L+2f_T)L\right] \geq 0.$$

If we neglect the role of terms on $L$, i.e., $f(R, L, T) = f(R, T)$, the above energy conditions reduce to that in $f(R, T)$ gravity [38, 39]. Further, by neglecting the dependence on the trace of the energy-momentum tensor, we can have the energy conditions in $f(R)$ gravity, which are consistent with the results in the literature [40, 41, 42].

## 5 Constraining $f(R, L, T)$ gravity

In our present approach, the energy-condition inequalities can be used to place bounds in a given $f(R, L, T)$ function in the context of FLRW models. To investigate such bounds, we first note that the Ricci scalar and its derivatives for a spatially flat FLRW geometry can be expressed in terms of the Hubble parameter $H$, deceleration parameter $q$, jerk $j$ and snap $s$ [20, 43, 44], that is,

$$R = 6H^2(1-q), \quad (20)$$
$$\dot{R} = 6H^3(j-q-2), \quad (21)$$
$$\ddot{R} = 6H^4(s+q^2+8q+6), \quad (22)$$

such that $q = -\frac{1}{H^2}\frac{\ddot{a}}{a}$, $j = \frac{1}{H^3}\frac{\dddot{a}}{a}$ and $s = \frac{1}{H^4}\frac{\ddddot{a}}{a}$. Therefore, the energy conditions for the $f(R, L, T)$ gravity can be rewritten as

$$SEC: (\rho+3p) + \frac{1}{8\pi+\frac{1}{2}(f_L+2f_T)}\left[f-6H^2(1-q)f_R+\right.$$
$$18H^4\left(6f_{RRR}H^2(j-q-2)^2+f_{RR}(s+q^2+7q+j+4)\right)$$
$$\left. -(f_L+2f_T)L\right] \geq 0, \quad (23)$$

$$NEC: (\rho+p) + \frac{6H^4}{8\pi+\frac{1}{2}(f_L+2f_T)}\left[6f_{RRR}H^2(j-q-2)^2\right.$$
$$\left. +f_{RR}(s+q^2+8q+6)\right] \geq 0, \quad (24)$$

$$DEC: (\rho-p) + \frac{1}{8\pi+\frac{1}{2}(f_L+2f_T)}\left[(f_L+2f_T)L-f+\right.$$
$$6H^2(1-q)f_R - 36H^6 f_{RRR}(j-q-2)^2 \quad (25)$$
$$\left. -6H^4 f_{RR}(s+q^2+3q+5j-4)\right] \geq 0,$$

$$WEC: \rho + \frac{1}{8\pi+\frac{1}{2}(f_L+2f_T)}\left[\frac{1}{2}(f_L+2f_T)L\right. \quad (26)$$
$$\left. -\frac{1}{2}(f-6H^2(1-q)f_R) - 18H^4 f_{RR}(j-q-2)\right] \geq 0.$$

To exemplify how the above conditions can be used to place bounds on $f(R, L, T)$ theories, we first note that, apart from the WEC, all of the above inequalities depend on the current value of the snap parameter. Therefore, since no reliable measurement of this parameter has been reported hitherto, in what follows, we shall focus on the WEC requirement in the confrontation of the energy-condition bounds with observational data.

We can consider $q_0 = -0.58$, $H_0 = 68.8$ and $j_0 = 1.15$, and $s_0 = -0.25$ according to References [45] to obtain the constraints or bounds on the model parameters for the present universe. For the cases considered in this work, we do not need these values for the WEC fulfilment, whereas the values are utilized to obtain the bounds for SEC violation.

### 5.1 $f(R, L, T) = R + \gamma T + \lambda L$ with $L = -\rho$

As a first case, we consider the simplest linear functional form of $f(R, L, T)$, namely $f(R, L, T) = R + \gamma T + \lambda L$, where $\gamma$ and $\lambda$ are free parameters. For this case, the WEC fulfilment condition can be written as

$$\rho - \frac{\gamma(\rho+3p)}{16\pi+\lambda+2\gamma} \geq 0. \quad (27)$$

The condition ensuring attractive gravity within this model is expressed as $8\pi + \frac{\lambda}{2} + \gamma > 0$ (derived from (12)). The validity of WEC reduces to $-\frac{\gamma}{2}(\rho+3p) \geq 0$, as it mandates that the energy density measured by any observer remains non-negative. The latter condition imposes a constraint on the parameter $\gamma$.

We outline constraints on the model parameters for the present universe, that is, the universe dominated by dark energy. We consider the universe characterized by $\omega \simeq -1$ and deduce further insights from the inequality (27), namely

$$\lambda > -16\pi - 2\gamma \quad \text{and} \quad \gamma \geq 0. \tag{28}$$

From Eq.(14), it becomes evident that achieving an accelerated expansion necessitates the condition $\rho_{eff} + 3p_{eff} < 0$. The conditions governing late-time accelerated expansion impose specific constraints on model parameters. The constraints are $\gamma > 0$ and $\lambda > -16\pi - 2\gamma$. It is worth noting that one can ensure that ordinary matter adheres to all the energy conditions while achieves acceleration for some suitable functional form of $f(R, L, T)$.

**5.2 $f(R, L, T) = R^n + \alpha L T + \beta$ with $L = -\rho$**

As a second example, we shall consider the form $f(R, L, T) = R^n + \alpha L T + \beta$, in which $n$ is an integer and $\alpha$ and $\beta$ are constants. This functional form has been proposed by the authors of the theory themselves [23]. For $n = 1$, it has a de-Sitter type evolutionary phase triggered by an extra term appearing as a novelty of the theory.

The inequality for the WEC fulfilment condition can be written in terms of the $q$, $j$ and $H$ parameters as

$$\rho + \frac{1}{8\pi + \frac{3\alpha}{2}(p-\rho)} \left[ \frac{1}{2} \left( 2\alpha\rho^2 - \beta + \right. \right. \tag{29}$$
$$\left. \left. \frac{6^n(1-n)\left(H^2(1-q)\right)^n \left(n(j-q-2) - (1-q)^2\right)}{(1-q)^2} \right) \right] \geq 0.$$

For simplicity, we consider the case $n = 1$, so that Equation (29) reduces to

$$\rho + \frac{2\alpha\rho^2 - \beta}{16\pi + 3\alpha(p-\rho)} \geq 0. \tag{30}$$

Since the term $16\pi + 3\alpha(p - \rho) > 0$ using (12), we should have $2\alpha\rho^2 - \beta \geq 0$. It is noted that WEC depends upon the sign of $\alpha$. Below, we present the energy conditions constraints for the model parameters at the present epoch, that is, the universe dominated by dark energy.

We again consider the universe characterized by $\omega \simeq -1$. From (30), WEC is obeyed when:

$$\alpha < \frac{8\pi}{3\rho} \quad \text{and} \quad \beta \leq 2\alpha\rho^2. \tag{31}$$

Likewise, within this model, it is evident that achieving an accelerated expansion necessitates violation of SEC. Consequently, one can aim to establish the bounds for specific values of $n$. Initially, we can fix the value of $n$ and obtain the constraints on $\alpha$ and $\beta$ for SEC violation (while WEC is satisfied). For $n = 1$, we have similar bounds as WEC but with the strict inequality for $\beta$. Given that the model incorporates higher powers of $R$ and the terms of $L$ and $T$, the expression for SEC would entail higher orders of $\rho$, $p$ and $n$. This allows for the determination of constraints for different values of $n$, which result in various classes of $f(R, L, T)$ models.

## 6 Summary and Discussion

We have worked with a generalized theory of gravity featuring a flexible coupling of geometry and matter. The gravitational lagrangian is derived by incorporating an arbitrary function of $f(R, L, T)$ into the Einstein-Hilbert action. This is basically an extension and generalization of two classes of gravitational theories with geometry-matter coupling, that is, $f(R, L)$ and $f(R, T)$ theories.

The flexibility inherent in the Lagrangian formulation raises the question of placing constraints on such a theory based on physical considerations. We have addressed this issue by establishing constraints on general and specific forms of $f(R, L, T)$ gravity. We accomplish this by examining the energy conditions. By utilizing the Raychaudhuri equation and imposing the condition of attractive gravity, we derived the energy conditions and focused on the WEC to obtain the constraints on model parameters. In a formal sense, the WEC, NEC, DEC and SEC can be expressed similarly to how they are in GR. We have explored the present universe, characterized by the equation of state $\omega \simeq -1$, aiming to establish constraints on the free parameters for two models.

The model parameters have a significant impact on the validation of the energy conditions. We have found that both the considered models satisfy the energy conditions, specifically the WEC, for the appropriate choices of free parameters. These results could be a powerful tool for probing the dark sector of the universe, as any deviations from standard GR could be highlighted.

The first model we have worked out, namely $f(R, L, T) = R + \gamma T + \lambda L$ was strongly constrained. Remarkably, in order to be in accordance with an accelerated expanding universe, $\gamma > 0$ when $\lambda > -16\pi - 2\gamma$. This is a constraint that must be satisfied when this model is confronted with cosmological observational data. Analogously, in the second model, the values for $\alpha$ and $\beta$ to be obtained from confrontation with observational data must be smaller than the maximum values here derived.

## Data availability

There are no new data associated with this article.

## Acknowledgements

PKS acknowledges the Science and Engineering Research Board, Department of Science and Technology,

Government of India, for financial support to carry out the Research project No.: CRG/2022/001847. We are very much grateful to the honorable referee and to the editor for the illuminating suggestions that have significantly improved our work in terms of research quality, and presentation.